\documentclass{amsproc}

\pdfoutput=1

\usepackage{epsfig,amsfonts}
\usepackage{amssymb}

\def\be{\begin{equation}}
\def\ee{\end{equation}}

\def\bea{\begin{eqnarray}}
\def\eea{\end{eqnarray}}
\def\e{\epsilon}

\def\a{\alpha}
\def\b{\beta}

\theoremstyle{definition}

\theoremstyle{remark}

\numberwithin{equation}{section}

\begin{document}

\title{Bethe Ansatz approach to quench dynamics in the Richardson model}

\author{Alexandre Faribault}
\address{Physics Department, Arnold Sommerfeld Center for Theoretical Physics,
and Center for NanoScience, \\
Ludwig-Maximilians-Universit\"at, Theresienstrasse 37, 80333 Munich, Germany}
\email{alexandre.faribault@physik.uni-muenchen.de}
\thanks{A.F. is supported by the DFG through SFB631, SFB-TR12 and the
Excellence Cluster "Nanosystems Initiative Munich (NIM)"}

\author{Pasquale Calabrese}
\address{Dipartimento di Fisica, 'Universit\`a di Pisa and  
INFN,
56127 Pisa, Italy}
\email{calabres@df.unipi.it}
\thanks{P.C. benefitted of a travel grant from ESF (INSTANS activity)}

\author{Jean-S\'{e}bastien Caux}
\address{Institute for Theoretical Physics, Universiteit van  
Amsterdam,
1018 XE Amsterdam, The Netherlands}
\email{J.S.Caux@uva.nl}
\thanks{All the authors are thankful for support from Stichting voor Fundamenteel Onderzoek der Materie (FOM) in the Netherlands }

\date{March 30, 2009.}


\begin{abstract}

By instantaneously changing a global parameter in an extended quantum system, an initially equilibrated state will afterwards undergo a complex non-equilibrium unitary evolution whose description is  extremely challenging. A non-perturbative method giving a controlled error in the long time limit remained highly desirable to understand general features of the quench induced quantum dynamics. In this paper we show how integrability (via the algebraic Bethe ansatz) gives one numerical access, in a nearly exact manner, to the dynamics resulting from a global interaction quench of an ensemble of fermions with pairing interactions (Richardson's model). This possibility is deeply linked to the specific structure of this particular integrable model which gives simple expressions for the scalar product of eigenstates of two different Hamiltonians. We show how, despite the fact that a sudden quench can create excitations at any frequency,  a drastic truncation of the Hilbert space can be carried out therefore allowing access to large systems. The small truncation error which results does not change with time and consequently the method grants access to a controlled description of the long time behavior which is a hard to reach limit with other numerical approaches. 

\end{abstract}

\maketitle

\section{Statement of the problem}

Due to fast dissipation rates, the unitary evolution of out-of-equilibrium quantum systems remained for a long time a question which remained mostly of academic interest. However, recent developments in the field of cold atomic systems \cite{exp}, in which tunability of the Hamiltonian combined to very low dissipation made this problem experimentally accessible and created a strong interest  \cite{th-mix,cc-06,gdlp-07,IS} in trying to draw a general theoretical understanding of far from equilibrium quantum dynamics. The interaction quench is a simple way to create a prepared out-of-equilibrium quantum state of the system. It is realized by supposing a quantum system is initially in an eigenstate (the ground state) of Hamiltonian $H_{g_0}$. Here $g_0$ is some global tunable parameter of the Hamiltonian. At time $t=0$, the parameter $g_0$ is instantaneously changed to $g$ and as a consequence the initial state $|\psi_{g_0}^0\rangle$ is no longer an eigenstate of  $H_{g}$. This process puts the system in a far from equilibrium state, whose subsequent evolution is fully determined by the new Hamiltonian $H_g$ and the initial conditions.

 The resulting dynamics of the wave function are easily formulated using elementary quantum mechanics since they simply are given by solving the time dependent Schr\"{o}dinger equation:
 
 \bea
 | \psi(t)\rangle=e^{-iH_gt}|\psi_{g_0}^\mu\rangle.
 \eea

Naturally, the evolution is greatly simplified by projecting the system onto the eigenbasis of $H_g$ defined by a set of eigenvectors $|\psi^\nu_g\rangle$ with eigenvalues $\omega^\nu$.

 \bea
 |\psi(t)\rangle=\sum_\nu e^{-i\omega^\nu t} \langle\psi^\nu_g |\psi_{g_0}^\mu\rangle |\psi_{g}^\nu\rangle.
 \eea

The resulting time-evolved average of any given operator $\mathcal{O}$ can also be written as a sum over form factors weighted by the projections of the initial ground-state onto the eigenbasis of the final Hamiltonian:

\bea
\langle\psi(t) |\mathcal{O} |\psi(t)\rangle=\sum_{\nu,\nu'} e^{-i(\omega^\nu-\omega^{\nu'}) t} \left[\langle\psi^\nu_g |\psi_{g_0}^\mu\rangle \langle\psi_{g_0}^\mu| \psi^{\nu'}_g \rangle \right]\langle\psi^{\nu'}_g |\mathcal{O}  |\psi_{g}^\nu\rangle.
\label{evolO}
 \eea

Not only does one need to be able to compute eigenstates and eigenenergies of both Hamiltonians, the exact treatment also requires the capacity to compute overlaps between the eigenstates of two different Hamiltonians. We will show how this apparent hurdle is simply dealt with, when using the results from the algebraic Bethe ansatz (ABA) within the fermionic pairing model treated here.

Moreover, as is readily seen, obtaining exact results necessitates a double sum over the full Hilbert space. Since its dimension generally grows factorially with system size, this is apparently heavily limiting as far as reachable system sizes is concerned. In the coming sections, we also show how the quench populates predominantly a small subset of the Hilbert space therefore allowing a drastic truncation with minimal loss of information ultimately granting access to systems much larger than one could naively expect. 

Most of the results presented here were previously published in \cite{fcc-09}. This paper however supplements this reference with additional details on the truncation scheme and also presents novel ideas on the relationship between weak and strong coupling eigenstates of this particular system.


\section{System}

We treat a model of spin $1/2$ fermions in a set of single particle energy levels which can accommodate up to two electrons (of opposite spin). Electrons interact through a typical  Bardeen-Cooper-Schrieffer (BCS) uniform s-wave pairing term which retains coupling only between time-reversed states:

\be
H = \sum_\a\sum_{\sigma} \frac{\e_\a}{2} c^{\dagger}_{\a \sigma}
c_{\a \sigma} - g \sum_{\a, \beta} c^{\dagger}_{\a+}c^{\dagger}_{\a-}
c_{\beta-} c_{\beta+}\,.
\label{eq:H_fermions}
\ee

The model is basically a discrete version of the celebrated BCS Hamiltonian to which it reduces in the thermodynamic limit \cite{bcs-57}. An Anderson pseudo-spin representation \cite{anderson} also exists by defining $S^-_\alpha=c_{\alpha,\downarrow}c_{\alpha,\uparrow}$, which leads to

\bea
H_{BCS}(\{\e_j\},g) = \displaystyle\sum_{j=1}^{N} \e_j S^z_j - g \sum_{\alpha,\beta=1}^{N} S_\a^+S_\b^-.
\label{HBCS}
\eea

Here the $N$ levels are the unblocked levels, i.e. the ones which are not occupied by a single electron and therefore participate to the dynamics  \cite{rs-62}. This model, although initially developed by Richardson \cite{rs-62} to describe pairing in nuclei, has also been used more recently to describe the physics of ultrasmall superconducting metallic grains \cite{rev}.
 
Irrespective of the distribution of the energy levels and of their degeneracies, the model is integrable \cite{camb} and can be diagonalized fully using the ABA.

\subsection{The eigenstates of the systems}

Through the use of the algebraic Bethe ansatz \cite{rev,BetheZP71,zlmg-02}, one can access a very compact representation of the eigenstates of Hamiltonian \ref{HBCS}. For any given set of $M$ complex parameters $w_j$ we shall call rapidities, the repeated action of the $C(w_k)$ operator of the monodromy matrix on a fully down polarized state, builds a "general" state:

\bea
|\{w_j\}\rangle = \prod_{k=1}^{M} \mathcal{C}(w_k) |\downarrow \downarrow \downarrow ...\downarrow \rangle \equiv \prod_{k=1}^{M}\left(\sum_{\a=1}^N \frac{S_\a^+}{w_k - \e_\a}\right)|\downarrow \downarrow \downarrow ...\downarrow \rangle\,.
\label{GENSTATE}
\eea

All of these $M$ rapidities states have a definite $z$-projection of the total spin operator which is simply given by $M-\frac{N}{2}$. For any value of $g$, the unnormalized eigenstates of the BCS Hamiltonian are all built in this fashion. For a given coupling constant $g$, the complete set of eigenstates correspond to the general states constructed out every set of rapidities which are solution to the Richardson (Bethe) non-linear algebraic system of equations:

\bea
-\frac1g=\sum_{\a=1}^N\frac1{w_j - \e_\a}-\sum_{k\neq j}^{M}  
\frac2{w_j-w_k}\,\quad
j=1,\dots, M\,.
\label{RICHEQ}
\eea

The eigenenergy of one of these states is given (up to a constant) by the sum
\be
E(\{w_j\}) = \sum_{j=1}^M w_j.
\label{energy}
\ee

At zero coupling, the full set of fixed magnetization (fixed $M$) eigenstates of Hamiltonian $H_{0}(\{\e_j\}) = \displaystyle\sum_{j=1}^{N} \e_j S^z_j$ is known to be given by the various colinear spins states with $M$ spins aligned in the $+\hat{z}$ direction and the remaining ones in the $-\hat{z}$ direction.  Each and every one of these states can be expressed in the form \ref{GENSTATE} by using a set of $M$ real rapidities $w_j$. By giving the $M$ rapidities the values of the $M$ energy levels at which a $+\frac{1}{2}$ spin is found ($\{w_j\} = \{e^*_j\}$), this leads to 

\be
|\{w_j\}\rangle \propto \left(\prod_{\{\e^*_i\}}^M S_i^+\right)|\downarrow \downarrow \downarrow ...\downarrow \rangle\,,
\ee

\noindent which is just the state with $M$ flipped spins at the chosen levels $\{\e^*_i\}$.

In the opposite regime of infinitely large coupling, the Hamiltonian reduces to 

\be
H_{\infty}(\{\e_j\}) = \displaystyle -g \sum_{\alpha,\beta=1}^{N} S_\a^+S_\b^-=-g S^+_{tot}S^-_{tot}=-g\left[ \boldsymbol{J}^2-(J^z)^2+J^z\right],
\ee

\noindent where $\boldsymbol{J}$ is the total angular moment operator of the $N$ spins system and $J^z$ is its z component. The eigenvectors of the Hamiltonian are therefore known through the procedure of angular momentum coupling of $N$ spins $\frac12$ leading to sets of eigenstates of $\boldsymbol{J}^2$ and $J_z$. As mentioned previously, the $z$-projection of the total angular momentum is equal to the number of rapidities $M$. For a fixed value of $M$, one can absorb the $M$-dependent terms in a global shift of the energies and consequently the eigenspectrum consists of $\frac{N!}{(N-M)!(M)!}$ states subdivided into degenerate sub-bands of total angular momentum $j$ with eigenenergies, and degeneracies given by:
\bea
E(j) &=& -g\left[j\left(j+1\right)\right]
\label{ENERGINF}
\\
D(j) &=& \frac{N!(2j+1)}{(\frac{N}{2}+j+1)!(\frac{N}{2}-j)!} 
\eea

In the strong coupling limit, the solutions to Richardson's equations naturally defines
 an alternative representation for these eigenstates. It is known that for a state defined by $M$ rapidities, the various solutions will be defined by a number $r$ of divergent rapidities. These rapidities can be expressed in the large $g$ limit as $w_i=L_i g$, where the $L_i$s are the zeros of the Laguerre polynomial \cite{largealt}:

\bea
L_r^{-1-(N+2r-2M)}(L_i)=0   \ \ \forall \ L_i.
\label{laguerre}
\eea

There is a direct correspondence between the number $r$ of diverging rapidities and the total spin eigenvalue $j$. Looking simply at the energies, it was shown \cite{largealt} that

\bea
r=j+M-\frac{N}{2}
\eea

The remaining roots (which remain finite) defining the various eigenstates can then be found by solving the following set of simplified Richardson's equations which involve only the non-divergent $M-r$ rapidities:

\bea
\sum_{\a=1}^N\frac1{w_j - \e_\a}-\sum_{k\neq j}^{M-r}  
\frac2{w_j-w_k} = 0\,\quad
j=1,\dots, M-r\,.
\label{RICHEQ_inf}
\eea

This particular set of equations still requires a numerical approach in order to find every possible solution.

One must realize that within a given degenerate subset both sets of eigenstates (Bethe Ansatz and total spin eigenbasis) are chosen differently. This is of course not an issue in the $g=\infty$ case, but the Bethe ansatz representation compares advantageously to the total spin eigenbasis when trying to expand in powers of $\frac{1}{g}$. Whereas in the former, one can directly expand the Richardson equations in powers of $\frac{1}{g}$ in order to describe the system \cite{largealt}, in the latter representation we would need to apply perturbation theory on a set of highly degenerate levels which would therefore involve diagonalization of large matrices.

\subsection{Numerical solution}
\label{nums}

Except in the previously discussed extreme cases of zero or infinite coupling, no analytical solution is known to the Richardson equations and one therefore needs to resort to numerical techniques in order to study this particular system.  Many efforts have been made to construct efficient algorithms for solving Richardson's equations  \cite{algo,largeroman} but it still remains a difficult task to achieve stable and fast computation of the various solutions. As is always the case when dealing with non-linear systems of algebraic equations, the convergence of any algorithm towards a given solution highly depends on the capacity one has to generate an appropriate initial approximation of it. 

In the absence of any known general features of the solutions in the intermediate coupling regime, it makes it quite natural to use a scanning procedure starting from the known solutions at $g=0$ and increasing the coupling in small steps in order to reach solutions at a given value of $g$. This guarantees an adequate trial solution from which an iterative procedure can be carried out successfully.

The solutions are such that rapidities are either real or form a complex conjugate pair (CCP) with another one. At the critical values of $g$ at which two real rapidities collapse into a CCP (or vice-versa), Richardson's equations expressed in the form \ref{RICHEQ} have cancelling divergent terms which require appropriate modifications in order to be dealt with numerically. These forming and splitting apart of CCPs happen exclusively when two rapidities, say $w_1$ and $w_2$ are equal to one another while at the same time being equal to one of the energy levels $\e_c \in \{\e_1,\e_2 ...\e_N\}$. This allows a rewriting of the equations using the two real variables:

\bea
\lambda_+&=&2\e_c-w_1-w_2
\nonumber\\
\lambda_-&=&(w_1-w_2)^2.
\label{CHANGEVAR}
\eea

Contrarily to $w_1$ and $w_2$, these variables maintain a finite derivative with respect to $g$ while going from real to complex. One can additionally transform the Richardson equations for $w_1$ and $w_2$ in the following two equations:
\bea
4\lambda_+ + \left((\lambda_+)^2-\lambda_-\right) G_1 = 0
\\
\left((\lambda_+)^2-\lambda_-\right) \lambda_- G_2 - 4\lambda_+=0
\eea

\noindent where

\bea
\ \ \ \ \ \ G_1=-\frac{2}{g} - \sum_{\a\ne c}^N\left[\frac1{w_1- \e_\a}+\frac1{w_2- \e_\a}\right]+2\sum_{k\neq 1,2}^{M}  
\left[\frac1{w_1-w_k}+\frac1{w_2-w_k}\right]
\\
G_2=-\sum_{\a\ne c}^N\left[\frac1{(w_1- \e_\a)(w_2- \e_\a)}\right]+2\sum_{k\neq 1,2}^{M}  
\left[\frac1{(w_1-w_k)(w_2-w_k)}\right]
\eea

Rapidities which form a CCP at a given $g$ can split apart into two real rapidities at a larger coupling and can also reform a CCP with a different rapidity.This means that one might need to redefine a new set of variables (and equations), at every step in $g$. This is simply achieved by "pairing" any two real rapidities $w_i,w_j$ which have the same closest single particle energy level $\e_c$, and using for them the variables \ref{CHANGEVAR}. For these rapidities we also use the modified version of the Richardson equations written above. This always results in a system of $M$ equations depending on $M$ variables, but it can be modified at every step in $g$ in order to adapt to the local structure of the solutions. 

Independently of the pairing used at a given point, the Jacobian is always obtainable analytically in terms of these variables and it therefore allows a straightforward use of Newton's algorithm for non-linear algebraic equations. Despite the lack of guaranteed stability (unavoidable in a general set of algebraic equations), one can still manage to compute rapidly any solution in this way, provided we adjust the steps in $g$ in such a way that at every $g$ we correctly evaluate the rapidities which need to be paired. In a case where numerous pairings-unpairings happen on a small $g$ interval, one simply needs to reduce the steps in $g$ in order to make sure every rapidity crosses only one critical point per step. Since the scanning process gives access to a local evaluation of the derivatives of rapidities, we can always get a good evaluation of the desired $\Delta g$ for the coming step.  

\begin{figure}[ht]
  \includegraphics[width=10.5cm]{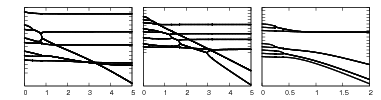}
\caption{A few examples of numerical solutions for M=8,N=16 (real part only) showing how to relate the $g=0$ structure to the one at $g=\infty$. We respectively get 2,3 and 5 divergent rapidities}
\label{examples}
\end{figure}

Figure \ref{examples} shows the real part of rapidities for a few solutions of the Richardson equations (examples showing also the imaginary part can be seen in other papers \cite{fcc-08} for example). These examples show how one can conjecture, for a given $g=0$ solution, the number of diverging rapidities that this state will have when deformed from $0$ to infinite coupling.

One can understand the passage of a given state from its uncoupled form to the resulting large coupling state, by looking at the contiguous blocks of occupied and unoccupied states at $g=0$. Any contiguous block of $m$ unoccupied states will refrain up to $m$ rapidities above it from diverging to infinity. Starting from high energies, one looks at the first block of occupied states, and the first following block of empty states. Supposing they respectively contain $p_1$ states and $h_1$ states, if $h_1 \ge p_1$, the $p_1$ rapidities will remain finite. If $h_1 < p_1$ then $h_1$ of these rapidities will remain finite and the remaining $p_1-h_1$ can be added to the following block of occupied states. Supposing the next blocks of occupied (empty) states contain $p_2$ ($h_2$) states, it would be treated in the same way, but using $p'_2 = \begin{cases}p_2\\p_2+p_1-h_1\end{cases}$ depending on what happened in the previous block. In the three examples presented in Figure \ref{quenchmat}, we respectively find for the leftmost case: $(p_1=1,h_1=2);(p_2=2,h_2=1);(p_3=2,h_3=1);(p_4=3,h_4=3)$ leading to two diverging rapidities (here they form a CCP), in the middle panel we have  $(p_1=3,h_1=1);(p_2=2,h_2=1);(p_3=1,h_3=1);(p_4=2,h_4=2)$ giving 3 diverging rapidities (two of them forming a CCP) and the last case has $(p_1=3,h_1=3);(p_2=5,h_2=0)$ giving 5 divergent rapidities (including 2 CCPs). Although giving results equivalent to the algorithm proposed in \cite{largeroman}, this approach can also provide some insight on approximative values of the finite rapidities in the large $g$ limit. This could give access, in this limit, to initial guesses close enough to the real solutions to allow direct computation of the various solutions to eq. \ref{RICHEQ_inf}.

This procedure seems to work well for any state which is half-filled or less, since in that case the diverging rapidities all have real parts going towards $-\infty$, the Laguerre polynomials (\ref{laguerre}) involved having all their zeros in the left half plane. At higher filling factors this is no longer true, but the alternative choice of representing the eigenstates as the repeated application of the operator $B(w_i)$ on the fully up-polarized pseudo-vacuum: $\prod_{k=1}^{M}\left(\sum_{\a=1}^N \frac{S_\a^-}{w_k - \e_\a}\right)|\uparrow \uparrow \uparrow ...\uparrow \rangle$, could probably allow similar arguments to be used in this case.

\section{Quench Matrix}

Having shown how one can access individually any eigenstate of the system, we now need to be able to realize explicit calculations of the projection of the initial state of the system onto the eigenbasis of the final Hamiltonian. For a general Hamiltonian and even for a general integrable Hamiltonian this problem requires full diagonalization, in a given basis, of the initial and final Hamiltonians. The factorial growth of the Hilbert space dimension (and therefore of the Hamiltonian matrix) limits one to very small system sizes.

For ABA integrable systems there exists a formula, due to Slavnov \cite{s-89}, which gives an expression of the overlap between an eigenstate of the system and a state built identically (eq. \ref{GENSTATE}) but for a general set of rapidities. This projection can be written as the determinant of an $M$ by $M$ square matrix, which in the model discussed here reads explicitly \cite{zlmg-02}:

\bea
\ \ \ \ \ \ \ \langle\{w_j\}|\{v_j\}\rangle = \frac{1}{\sqrt{{\rm Det}G[\{w_j\}]{\rm Det}G[\{v_j\}]}}\frac{\displaystyle\prod_{a\ne b}^M v_b-w_a}{\displaystyle\prod_{a<b}^M(w_a-w_b)\displaystyle\prod_{a>b}^M(v_a-v_b)} {\rm Det} J\, ,
\label{SLAVNOV}
\eea

\noindent  with the matrix elements  of $J$ given by:

\bea
\ \ J_{ab} = \frac{v_b-w_b}{v_b-w_a} \left[\sum_{k=1}^N\frac{1}{(w_a-\e_k)(v_b-\e_k)} - 2 \sum_{c\ne a} \frac{1}{(w_a-w_c)(v_a-w_c)}\right].
\label{JAB}
\eea

The $M$ by $M$ Gaudin matrices $G$, whose presence is needed to normalize both eigenstates, are themselves obtained through the limit $\{w_j\} \to \{v_j\}$ of eq. \ref{SLAVNOV}, which gives:

\bea
G_{ab}[\{v_j\}] = 
\begin{cases}\displaystyle
\displaystyle\sum_{k=1}^N \frac1{(v_a-\e_k)^2}-2\displaystyle\sum_{c\neq  
a}^{M}\frac1{(v_a-v_c)^2}\quad &
a=b\,,\\ \displaystyle
\frac2{(v_a-v_b)^2}& a\neq b\,.
\end{cases}
\eea

In the precise case treated here, the $C(w)$ operators have no explicit dependency on $g$ (see eq. \ref{GENSTATE}). Consequently eigenstates of the Hamiltonian at any value of $g$ are all general states built \`{a} la eq. \ref{GENSTATE}. This allows one to use formula \ref{SLAVNOV} directly to compute the relevant overlaps between two sets of eigenstates $\langle\psi^\nu_{g} | \psi^\mu_{g_0}\rangle$. One must understand that a different model might not share this property and if a determinant representation exists for these overlaps in a general integrable model, it remains to be found.

\subsection{Weak coupling to strong coupling}

\begin{figure}[ht]
\includegraphics[width=8.5cm]{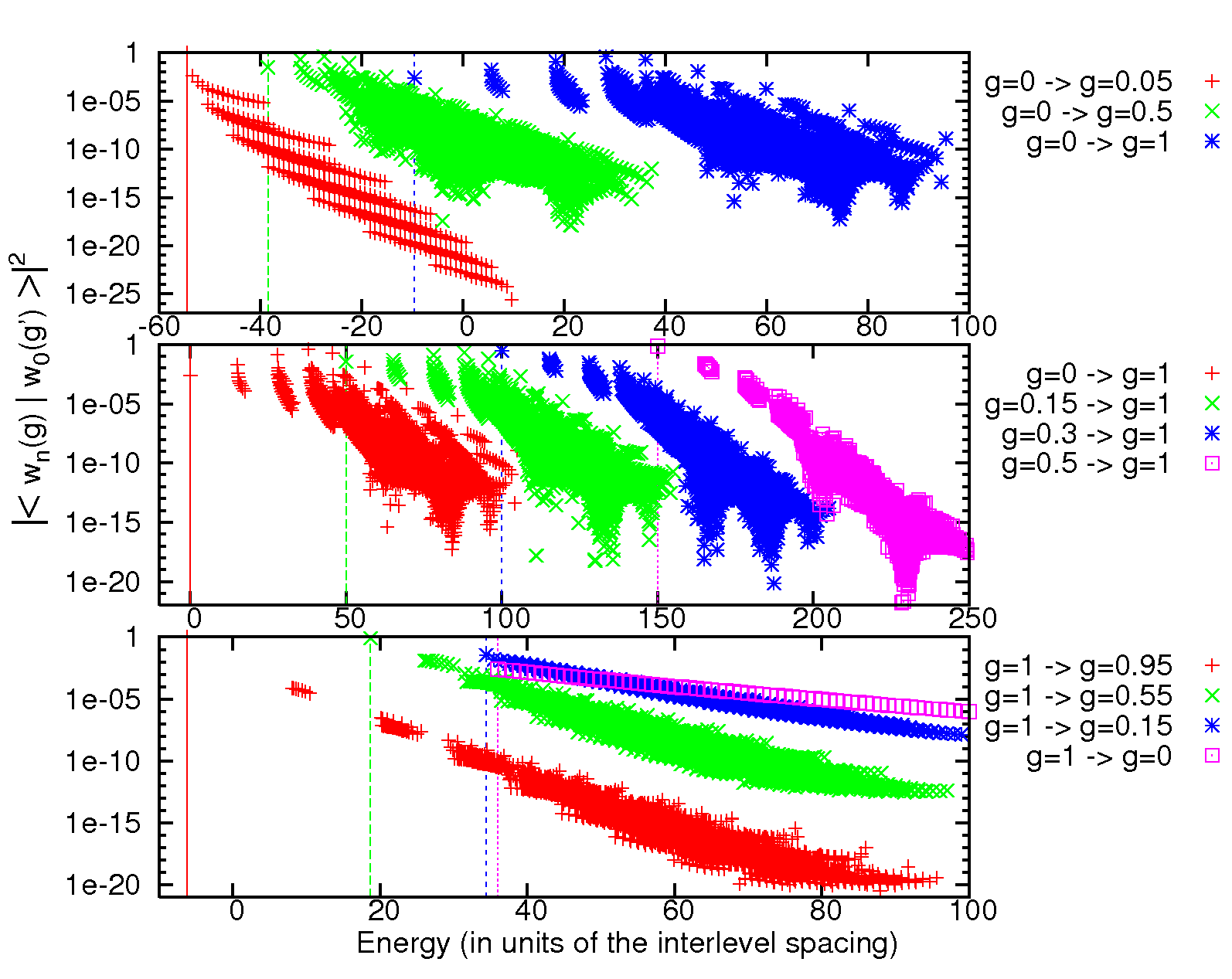} 
\caption{First column of the quench matrix (ground-state overlaps) for several
quenches. In all plots $N =16,M= 8$ and the ground state energies (represented
by vertical lines) have been shifted for clarity.
Top: Decomposition of the $g=0$ ground-state with states at $g=0.05,0.5,0.95$. 
Center: Decomposition of several initial ground-state $g_0=0,0.15,0.3,0.5$
in terms of the states at $g=1$.
Bottom: Decomposition of the $g_0=1$ ground-state in terms of 
$g=0.95,0.55,0.15, 0$ states (from ref. \cite{fcc-09}).
}
\label{quenchmat}
\end{figure}

For $M=8$ rapidities and $N=16$ energy level, the dimension of the Hilbert space is 12870 which is sufficiently small to allow us to compute the full set of eigenstates. Using Slavnov's formula \ref{SLAVNOV} we calculate the overlaps $\langle\psi^\nu_g|\psi^0_{g_0}\rangle$ of the $g_0$ ground state with the full eigenbasis at $g$. For a few quenches, focusing on equally spaced levels $(\epsilon_{i+1}-\epsilon_i=1)$, the squared norm of these overlaps are plotted on figure \ref{quenchmat} as a function of the energy of the state $|\psi^\nu_g\rangle$.

In the limit of a small quench from $g=0$ to $g=\Delta g$, first order perturbation theory theory can be used. It gives non-zero overlaps with the $g=\Delta g$ coupled ground state and the set of states which differed (at $g=0$) from the ground state by a single particle-hole excitation (one up spin moved to a different level). The following order would allow non-zero overlaps with two particle-hole excitations states and so on. This fact is clearly seen in the smallest quench presented in the top panel ($g=0.05$) of figure \ref{quenchmat} showing a set of decreasingly overlaping subbands, each associated with a different number of particle hole excitations

\begin{figure}[ht]
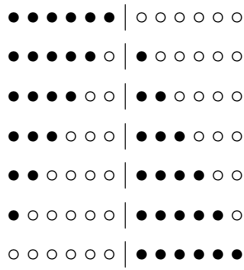

 \bea
\bullet\bullet\bullet\bullet\bullet\bullet | \circ\circ\circ\circ\circ\circ
\nonumber\\
\bullet\bullet\bullet\bullet\bullet\circ | \bullet\circ\circ\circ\circ\circ
\nonumber\\\bullet\bullet\bullet\bullet\circ\circ | \bullet\bullet\circ\circ\circ\circ\nonumber\\\bullet\bullet\bullet\circ\circ\circ | \bullet\bullet\bullet\circ\circ\circ
\nonumber\\
\bullet\bullet\circ\circ\circ\circ | \bullet\bullet\bullet\bullet\circ\circ
\nonumber\\
\bullet\circ\circ\circ\circ\circ | \bullet\bullet\bullet\bullet\bullet\circ
\nonumber\\
\circ\circ\circ\circ\circ\circ | \bullet\bullet\bullet\bullet\bullet\bullet
\nonumber
\eea
\caption{Pictorial representation of the "single block states" 
obtained by promoting contiguous blocks of $N_p \le M$ rapidities 
from right below to right above the Fermi level.}
\label{M00}
\end{figure}

When the quench is carried from an initial weak-coupling ground state to a stronger coupling (the two top panels of figure \ref{quenchmat}), one finds that a surprisingly small fraction of the final Hilbert space is importantly populated by the quench. In fact we see, in the distribution of energies, the large coupling band structure (see eq. \ref{ENERGINF}) forming and notice that in each of the bands the projection on a single state of the initial ground state heavily dominates. We could easily verify that these states are the finite $g$ deformations of a given set of $g=0$ configurations pictorially shown on figure \ref{M00}. This set of states is obtained (at $g=0$) by promoting contiguous blocks of $N_p \le M$ rapidities from right below to right above the Fermi level (FL). According to the procedure outlined in section \ref{nums}, they all lead, at $g\to\infty$, to states whose number of rapidities remaining finite is equal to $N_p$. In fact, they are the lowest energy states having a given number of particle-hole excitations (at $g=0$) equal to the number of finite rapidities (at $g \to \infty$).

\begin{figure}[ht]
  \includegraphics[width=4.5cm]{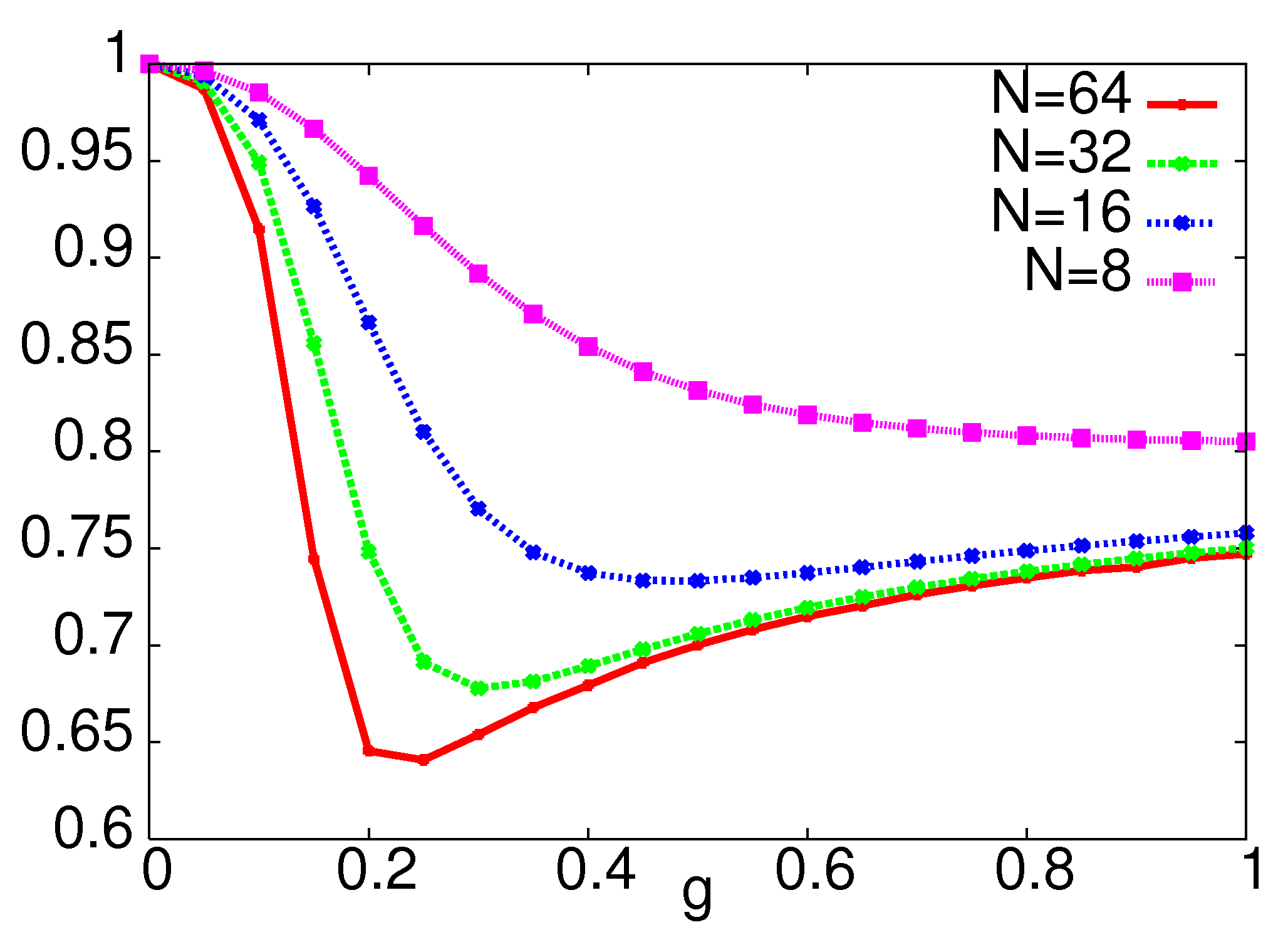}
\caption{Total contribution of these states 
to the amplitude of the initial state at $g_0=0$, as a function of 
interaction (from ref. \cite{fcc-09}).}
\label{M00contrib}
\end{figure}

For any quench from $g_0=0$ to $g\in[0,1]$, this set of $M+1$ states always account for more than 60\% of the total amplitude $\displaystyle \sum_{\nu} |\langle \psi^\nu_g | \psi^0\rangle|^2$ of the wave function expressed in the final eigenbasis. Figure \ref{M00contrib} shows this total contribution when quenching from an uncoupled initial state. Despite the presence of a dip in their contribution when quenching to the intermediate coupling regime, they clearly remain strongly dominant for finite size systems.

\begin{figure}[ht]
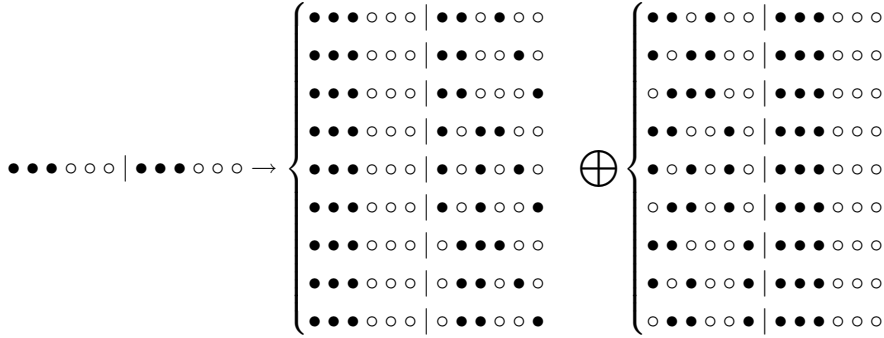

  \bea\nonumber
\bullet\bullet\bullet\circ\circ\circ | \bullet\bullet\bullet\circ\circ\circ \to \begin{cases}\bullet\bullet\bullet\circ\circ\circ | \bullet\bullet\circ\bullet\circ\circ
\\
\bullet\bullet\bullet\circ\circ\circ | \bullet\bullet\circ\circ\bullet\circ
\\
\bullet\bullet\bullet\circ\circ\circ | \bullet\bullet\circ\circ\circ\bullet
\\
\bullet\bullet\bullet\circ\circ\circ | \bullet\circ\bullet\bullet\circ\circ
\\
\bullet\bullet\bullet\circ\circ\circ | \bullet\circ\bullet\circ\bullet\circ
\\
\bullet\bullet\bullet\circ\circ\circ | \bullet\circ\bullet\circ\circ\bullet
\\
\bullet\bullet\bullet\circ\circ\circ |\circ \bullet\bullet\bullet\circ\circ
\\
\bullet\bullet\bullet\circ\circ\circ | \circ\bullet\bullet\circ\bullet\circ
\\
\bullet\bullet\bullet\circ\circ\circ | \circ\bullet\bullet\circ\circ\bullet
\end{cases}
\bigoplus
 \begin{cases}
\bullet\bullet\circ\bullet\circ\circ | \bullet\bullet\bullet\circ\circ\circ
\\
\bullet\circ\bullet\bullet\circ\circ | \bullet\bullet\bullet\circ\circ\circ
\\
\circ\bullet\bullet\bullet\circ\circ | \bullet\bullet\bullet\circ\circ\circ
\\
\bullet\bullet\circ\circ\bullet\circ | \bullet\bullet\bullet\circ\circ\circ
\\
\bullet\circ\bullet\circ\bullet\circ | \bullet\bullet\bullet\circ\circ\circ
\\
\circ\bullet\bullet\circ\bullet\circ | \bullet\bullet\bullet\circ\circ\circ
\\
\bullet\bullet\circ\circ\circ\bullet | \bullet\bullet\bullet\circ\circ\circ
\\
\bullet\circ\bullet\circ\circ\bullet | \bullet\bullet\bullet\circ\circ\circ
\\
\circ\bullet\bullet\circ\circ\bullet | \bullet\bullet\bullet\circ\circ\circ
\end{cases}
\eea
  \caption{Pictorial representation of the single-excitations for a given block state. One should understand that the same process is applied to every one of the block states (see Fig. \ref{M00})}
\label{SINGLE}
\end{figure}

It is then natural, when looking for additional large contributions, to assume that states built by slightly deforming the preceding set would be dominant. We therefore add to the reduced eigenbasis the set of states built at $g=0$ by adding a single particle-hole excitation to the block-states. Figure \ref{SINGLE}, shows, for one of the block states discussed previously, the $g=0$ configurations created by adding this single excitation (either above of below the Fermi level).

\begin{figure}[ht]
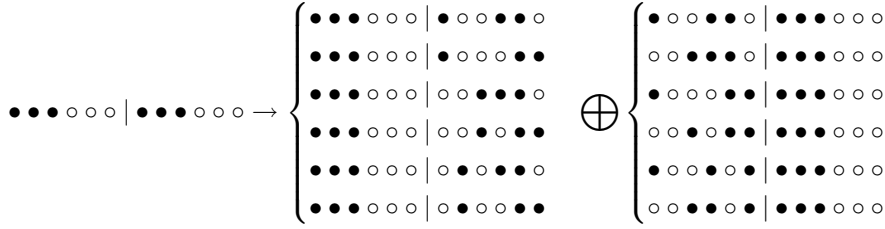

  \bea\nonumber
\bullet\bullet\bullet\circ\circ\circ | \bullet\bullet\bullet\circ\circ\circ \to \begin{cases}
\bullet\bullet\bullet\circ\circ\circ | \bullet\circ\circ\bullet\bullet\circ
\\
\bullet\bullet\bullet\circ\circ\circ | \bullet\circ\circ\circ\bullet\bullet
\\
\bullet\bullet\bullet\circ\circ\circ | \circ\circ\bullet\bullet\bullet\circ
\\
\bullet\bullet\bullet\circ\circ\circ | \circ\circ\bullet\circ\bullet\bullet
\\
\bullet\bullet\bullet\circ\circ\circ | \circ\bullet\circ\bullet\bullet\circ
\\
\bullet\bullet\bullet\circ\circ\circ | \circ\bullet\circ\circ\bullet\bullet
\end{cases}
\bigoplus
 \begin{cases}
\bullet\circ\circ\bullet\bullet\circ| \bullet\bullet\bullet\circ\circ\circ
\\
\circ\circ\bullet\bullet\bullet\circ| \bullet\bullet\bullet\circ\circ\circ
\\
\bullet\circ\circ\circ\bullet\bullet| \bullet\bullet\bullet\circ\circ\circ
\\
\circ\circ\bullet\circ\bullet\bullet| \bullet\bullet\bullet\circ\circ\circ
\\
\bullet\circ\circ\bullet\circ\bullet| \bullet\bullet\bullet\circ\circ\circ
\\
\circ\circ\bullet\bullet\circ\bullet| \bullet\bullet\bullet\circ\circ\circ
\end{cases}
\eea
  \caption{Pictorial representation of the doubly-excited states which are kept for a given block state. The same process is applied to every one of the block states (see Fig. \ref{M00}). Moreover, one should remember the inclusion of states featuring one excitation above and one below the FL (not represented here but discussed in the text).}
\label{DOUBLE}
\end{figure}

One can then add a set of states built with two particle-holes excitations over the block states. We include every state obtained with one particle excitation above the Fermi level and one hole excitation below it. These would be constructed by the tensor product of the possible states above the FL, as seen in the first set of states in Figure \ref{SINGLE}, with the possible states below the FL in the second set. Naturally, states with two excitations above (or below) the FL are also possible. A fairly small set of these is actually included, since the study of the quench matrix made for a smaller system indicates that they should be dominant. The treated cases are built by promoting any two rapidities (absence of rapidities) above (below) the Fermi level to any contiguous set of two free (occupied) level above the FL as shown in figure \ref{DOUBLE}.

In the case of $M=16, N=32$, the inclusion of all the aforementioned states proves sufficient to obtain more than 97\% of the weight of the wavefunction after the quench, i.e.: $\displaystyle \sum_{\nu} |\langle \psi^\nu_g | \psi^0\rangle|^2 \ge 0.97$. This figure holds for any final value of $g$ between $0$ and $1$. One must understand that for a given value of the final coupling only a part of these states will have an important contribution to the wavefunction. As one can readily see in Figure \ref{quenchmat}, the weight is shifted  to different parts of the Hilbert space as $g$ changes. For example, when doing a large quench the overlap of the final ground state with the initial one will become quite small, but for a small quench it would be the dominantly populated state. It is therefore possible to get an accurate projection of the uncoupled ground-state onto the coupled eigenbasis by keeping less than a 1000 states (out of the 601080390-dimensional Hilbert space for $M=16, N=32$) for a given quench. However, in order to be able to treat the spectrum of quenches discussed here we need to compute only a total of 7675 states. This drastic reduction of the effective Hilbert space is the key element which allows us to treat relatively larger system sizes with a minimal error.

Since this truncation scheme is carried out in the eigenbasis of the Hamiltonian driving the time evolution of the system, the error will not change with time. In fact, the time evolution only affects the phase of various components of the wave function and not their amplitude. Apart from the initial condition, the wavefunction (and by extension expectation value of operators) at any time does not depend on the previous history. For very large times, it can therefore be calculated directly and with a precision set by the small truncation error.

\subsection{Strong coupling to weak coupling}

Quenching form a strongly coupled initial ground state down to a weakly coupled system brings a radically different structure. Looking at figure \ref{quenchmat}, we see that although a small quench still brings a quench matrix naturally strongly peaked at low energies (with a roughly exponential decay in energy), when the difference between initial and final $g$ is made larger, the distribution flattens significantly.

One can understand this tendency by noting that the infinite coupling ground state is characterized by having every rapidity present diverging (see eq. \ref{energy}). In this case we have $\lim_{v\to\infty} C(v) \propto \sum_i S^+_i$. The repeated action of the operator for a set of diverging rapidities therefore make the $M$ rapidities ground state a uniform superposition of every $g=0$ $M$-flipped spins state, i.e. 
\bea
|\psi^0_{g\to\infty}\rangle \propto \left(\sum_iS_i^+\right)^M|\downarrow \downarrow \downarrow ...\downarrow \rangle.
\eea 
\noindent The $g\to\infty$ ground state therefore has a uniform projection onto the $g=0$ eigenbasis. In this extreme scenario, any truncation of the final Hilbert space is impossible. However, when starting from a finite albeit large initial coupling, it appears that the nearly monotonic energy behaviour could allow one to reduce the effective dimension of the problem by removing the largest energy states. The slow decrease of the overlaps in energy would still make the problem of very large to very weak $g$ quench hard to treat using this technique.

\section{Form factors}

Being in a position to compute the eigenstates, eigenvalues and the quench matrix elements is sufficient to access the time-evolved wave function of the system. However, to compute the average value of an observable $\mathcal{O}$, one still needs tractable expressions for the form factors:

\bea
\langle\psi^\nu_{g} | \mathcal{O}| \psi^\mu_{g}\rangle.
\eea

\noindent of the operator we are interested in. In this work we will look at time evolution of the off-diagonal order parameter defined as 

\bea
\Psi(t) = \frac{1}{M}\sum_{\a\b}\langle\psi(t)|S^+_\a S^-_\b|\psi(t)\rangle.
\label{OOP}
\eea

In equilibrium and in the thermodynamic limit, this quantity is closely related to the BCS gap making it a suitable parameter to quantify the superconducting tendency for finite systems even out of equilibrium. One could also use in a similar fashion the canonical order parameter $\sum_\a \sqrt{\frac{1}{4}-\langle S^z_\a\rangle^2}$ (as in \cite{bl-06,bcs}), but since both quantities showed the same qualitative behavior we only present the former here.

Although in the infinite coupling case we have $\displaystyle \sum_{\a\b}\langle\psi^\nu_{g} | S^+_\a S^-_\b | \psi^\mu_{g}\rangle\propto \delta_{\mu,\nu}$, since  $\displaystyle \sum_{\a\b}S^+_\a S^-_\b\propto H_\infty$, at any finite coupling it requires computation, for every couple $\a,\b$ of the form factors $\langle\psi^\nu_{g} | S^+_\a S^-_\b | \psi^\mu_{g}\rangle$.

Using Slavnov's formula, the solution to the inverse problem, i.e:
\bea
\displaystyle S^+_i&=&\lim_{u\to\e_i}(u-\e_i)C(u)\nonumber\\ S^-_i&=&\lim_{u\to\e_i}(u-\e_i)B(u), \eea
\noindent and determinant
properties, it was shown in \cite{fcc-09} that these form factors for $S^-_\a S^+_\b$ (and consequently $S^-_\a S^+_\b$ too) can be written as a sum of $M$ determinants. This was done by generalizing to the case $\{v\}\ne\{w\}$ the method of Ref. \cite{fcc-08}, starting from the double sums given in Ref. \cite{zlmg-02}.
For $\a \neq \b$ we have 
\begin{eqnarray}
 \langle \{v\}|S^-_\a S^+_\b|\{w\}\rangle &=&
\frac{\displaystyle\prod_{c}\left(\frac{v_c-\e_\a}{w_c-\e_\a}\right)\displaystyle\prod_{k\ne q}\left(w_k-w_q\right)}{
\displaystyle\prod_{b>a}(v_a-v_b)\prod_{a>b}(w_a-w_b)} \sum _{q=1}^{M} 
\frac{w_q-\e_\a}{w_q-\e_\b}
{\rm det} {\mathcal J}^q_{\alpha,\beta}\,,
\end{eqnarray}
where, defining $A_{ab}=J_{a b}\displaystyle\prod_{c\ne b} (v_c-w_b)$ ($J_{a b}$ being defined by eq. \ref{JAB}), the matrix elements are given by
{\small
\begin{eqnarray}
{\mathcal J}^q_{ab} &=&A_{a b}- \frac{\prod_{k\ne b,q}(w_k-w_b)}
{\prod_{k\ne b+1,q}(w_k-w_{b+1})} \frac{w_b-\e_\a}{w_{b+1}-\e_\a}
A_{a b+1},\ \ b<q-1,\nonumber\\
{\mathcal J}^q_{aq-1} &=&  A_{a q-1} +2\frac{(w_q-\e_\b)(w_{q-1}-\e_\a)}{
w_{q-1}-w_q}\prod_c\left(
\frac{v_c-\e_\b}{w_c-\e_\b} 
\right)
\nonumber\\&&\times
\prod_{k\ne q-1} (w_k-w_{q-1})
\frac{(2v_a-\e_\a-\e_\b)}{(v_a-\e_\a)^2(v_a-\e_\b)^2} 
,
\nonumber\\
{\mathcal J}^q_{aq} &=& 1/(v_a - \e_\a)^2, 
\nonumber\\
{\mathcal J}^q_{ab} &=&  A_{a b}, \ \  \ b > q\,.
\end{eqnarray}
}

For $\alpha=\beta$ it can be easily related to the form factors for $S^z_\a$ which were already published elsewhere \cite{zlmg-02}.

\section{Order parameter evolution}

\begin{figure}[ht]
\includegraphics[width=8.5cm]{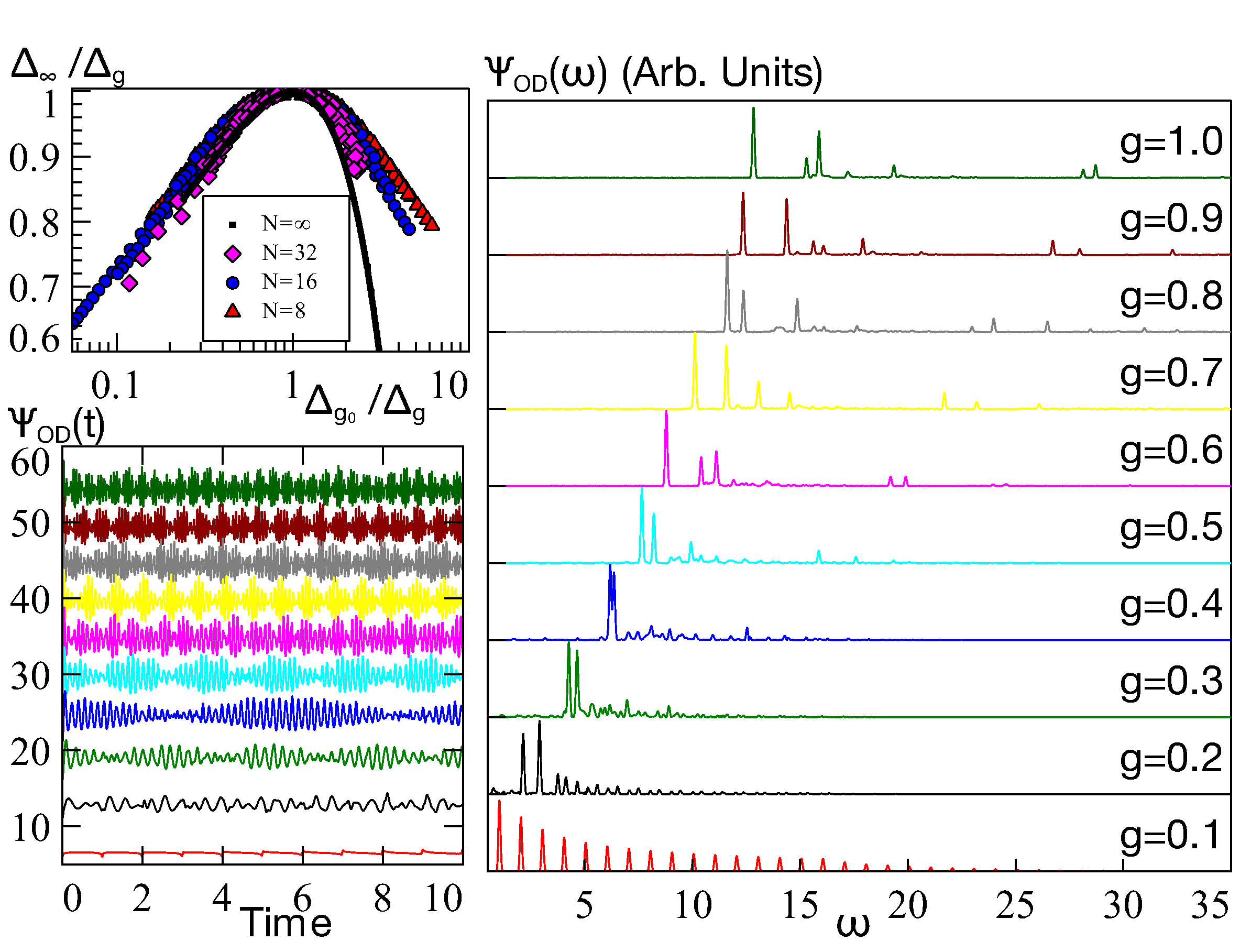}
\caption{Bottom left: Off-diagonal order parameter evolution for $N=32, M=16$. 
Right: Fourier transform,
the various plots are shifted on the vertical axis for clarity.
Top Left: Non-equilibrium finite-size ``phase diagram'' resulting from the 
time-averaged canonical gap obtained from the off-diagonal order parameter,
as explained in the text (from ref. \cite{fcc-09}).
}
\label{OP}
\end{figure}

Having a numerically tractable expression for every necessary element, is it straightforward to compute the time evolution of the off-diagonal order parameter (eq. \ref{OOP}). The upper left corner of  figure \ref{OP} compares the time-averaged value i.e.:

\be
\overline{\Psi}= \lim_{T\to\infty} \int_0^T dt \Psi(t) =
\sum_\nu |\langle\psi^\nu_g |\psi_{g_0}^0\rangle|^2 \langle\psi^\nu_g|\frac{1}{M}\sum_{\a,\b}S^+_\a S^-_\b|\psi^\nu_g\rangle
\ee

\noindent of the off diagonal order parameter with the thermodynamic limit given by the mean-field BCS theory. The non-equilibrium `phase diagram' for $N\to\infty$ shows that the final asymptotic value of the canonical gap $\Delta_\infty$ defines a universal curve when expressing $\Delta_\infty/\Delta_g$ 
vs $\Delta_{g_0}/\Delta_g$, where $\Delta_g$ is the equilibrium value (ground state expectation value)
at coupling $g$  \cite{bl-06}  .  In term of the off-diagonal order parameter, the gaps are given by $\Delta_g= g\sqrt{ (\langle\psi^0_g|\sum^N_{\a,\b=1}S^+_\a S^-_\b|\psi^0_g\rangle/M-1)}/M$ 
\cite{fcc-08} and we use $\Delta_\infty= g\sqrt{\left(\overline{\Psi}/M-1\right)}/M$. One can clearly see that for $\Delta_{g_0}/\Delta_g > 1$, which corresponds to a quench going from large coupling to weak coupling,  the results show stronger deviations from the BCS mean-field result obtained in the limit $N\to \infty$. This is to be expected, since for a finite system, at weak coupling, the superconducting correlations between electron pairs are suppressed when compared to the infinite system BCS result. In the limit $g\to0^+$, the infinite system still shows an instability which leads to a BCS wavefunction, but for any finite system size, the small coupling limit leads to a ground state formed by uncorrelated electron pairs and the mean-field BCS treatment loses validity. Nevertheless, in the regime of relatively small quenches ($\Delta_{g_0}/\Delta_g \sim 1$) one still sees that the impact of quantum fluctuations (which are not present in the BCS mean-field tratment) on the long time average gap is strongly suppressed even for the relatively small system sizes treated here.

The two additional panels in figure \ref{OP} show the time-evolution (and its Fourier transform) of the off-diagonal order parameter. Those results are given for $M=16,N=32$ which therefore necessitates the use of the previously described truncated eigenbasis. We focus on quenches from the $g=0$ uncorrelated ground-state to various finite values of the coupling.

For general quenches, the mean-field treatment, is known to lead to integrable classical dynamics, for which, in the steady-state reached for quenches from small to large coupling, the time evolution of the BCS gap is given by the Jacobi elliptic function \cite{bl-06,bcs}:

\bea
\Delta(t) = \Delta_+\mathrm{dn}[\Delta_+(t-\tau_0),k], \ \ \ \ k=1-\Delta^2_-/\Delta^2_+,
\eea

\noindent with parameters $\Delta_+ \sim \Delta_g$ and $\Delta_- \to 0$ for a weakly coupled $g\to 0^+$initial state. This gives rise to non-harmonic persistent oscillations which are periodic in time with a period given by the complete elliptic integral of the first kind:

\bea
T = \frac{2}{\Delta_+}\int_0^{\pi/2} \frac{d\phi}{\sqrt{1-k^2\sin^2\phi}} .
\eea

The mean-field solution should therefore show a Fourier transform made of equally spaced peaks which, as one can see, is quite different from the results obtained here. For small values of the final $g$, this difference is no surprise since the mean-field treatment assumes a BCS-like wave function, which, for a finite size system in sufficiently weak coupling is not realized. However, when $g \gtrsim g^*=(2\ln N)^{-1}$, it was shown \cite{fcc-08,ao-02} that the equilibrium static correlation functions are undistinguishable from the BCS correlations. For the largest final $g$ values shown here this inequality is respected but the quench dynamics are still qualitatively different from the expected BCS result.
 
One can understand this discrepancy by simply looking at the overlaps plotted on figure \ref{quenchmat}. It was clear from the start that an instantaneous quench can, in principle, create excitations at arbitrarily large frequencies. What our calculations show is that the excitation is done in a highly non-thermal way, by creating preferentially excitations in the mid-range energy spectrum. On the other hand, static correlations functions depend mostly on low-energy properties. Since the equispaced low energy bands which characterize the mean-field BCS regime are already well formed at $g \gtrsim g^*=(2\ln N)^{-1}$ (see the energy distribution on figure \ref{quenchmat}), it is therefore expected that low energy properties are quite similar to the BCS ones. The quench dynamics, however, by probing higher energy properties which still differ strongly from the BCS behaviour,  will retain a visible impact of the quantum fluctuations that the mean-field treatment cannot capture. As it was proposed in other models \cite{gdlp-07}, the experimental accessibility to quantum quenches could provide a powerful spectroscopic tool to study quantum fluctuations in the pairing properties of fermions. 

\section{Discussion}

In this work we have reported a new approach to the study the dynamics of a quantum system driven out of equilibrium by a global interaction quench. It uses integrability which provides numerically tractable ways to compute eigenstates, eigenenergies and form factors of various operators. We were able to exploit the peculiar property of the Richardson model in order to find simple expressions giving the overlaps of the eigenstates of the pre and post quench Hamiltonians. 

The obtained results show that the particular structure of these overlaps, also allows one to design a simple but very efficient truncated Hilbert space giving access to nearly exact results in systems too large to allow the full Hilbert space to be treated.

The generalization to other integrable models such as quantum spin chains \cite{cm-05} or atomic Bose gases \cite{cc-06b},  would be a highly desirable feat. However, in other general integrable models, the current lack of known tractable expressions for the elements of the quench matrix hinders such progress for now. Such a representation would allow exact calculation of quench behavior for a large variety of potential experiments.

\bibliographystyle{amsalpha}

\begin{thebibliography}{A}


\bibitem{exp} 
M. Greiner,   O.~Mandel, T.~W.~H\"ansch, \& I.~Bloch,
Collapse and revival of the matter wave field of a Bose-Einstein condensate,
Nature {\bf 419}, 51 (2002); \\
T. Kinoshita , T. Wenger, \& D. S. Weiss, 
A quantum Newton's cradle,
Nature {\bf 440}, 900 (2006); \\
L.E. Sadler, J. M. Higbie, S. R. Leslie, M. Vengalattore, 
\& D. M. Stamper-Kurn,
Spontaneous symmetry breaking in a quenched ferromagnetic spinor 
Bose-Einstein condensate,
Nature {\bf 443}, 312 (2006);\\ 
S. Hofferberth, I. Lesanovsky, B. Fischer, T. Schumm, \& J. Schmiedmayer,
Non-equilibrium coherence dynamics in one-dimensional Bose gases,
Nature {\bf 449}, 324 (2007);\\
C. N. Weiler, T. W. Neely, D. R. Scherer, A. S. Bradley, M. J. Davis \& 
B. P. Anderson, Spontaneous vortices in the formation of Bose-Einstein 
condensates, Nature {\bf 455}, 948 (2008).

\bibitem{th-mix} 
M. Cramer, C.M. Dawson, J. Eisert, \& T.J. Osborne,
Quenching, relaxation, and a central limit theorem for quantum lattice systems,
Phys. Rev. Lett. {\bf 100}, 030602 (2008);\\
C. Kollath, A. Laeuchli, \& E. Altman,
Quench dynamics and non equilibrium phase diagram of the Bose-Hubbard model,
Phys. Rev. Lett. {\bf 98}, 180601 (2007);\\
S. R. Manmana, S. Wessel, R. M. Noack, \& A. Muramatsu,
Strongly correlated fermions after a quantum quench,
Phys. Rev. Lett. 98, 210405 (2007);\\
M. Rigol, V. Dunjko, V. Yurovsky, \&  M. Olshanii,
Relaxation in a completely integrable many-body quantum system: An ab initio
study of the dynamics of the highly excited states of lattice hard-core bosons,
Phys. Rev. Lett. {\bf 98}, 050405 (2007);\\
M. A. Cazalilla, Effect of suddenly turning on interactions in the Luttinger model, Phys. Rev. Lett. {\bf 97} 156403 (2006);\\
T. Barthel \& U. Schollwock, 
Dephasing and the steady state in quantum many-particle systems
Phys. Rev. Lett. {\bf 100}, 100601 (2008);\\
M. Cramer, A. Flesch, I. P. McCulloch, U. Schollwoeck, \& J. Eisert,
Exploring local quantum many-body relaxation by atoms in optical superlattices,
Phys. Rev. Lett. {\bf 101}, 063001 (2008);\\
M. Rigol, V. Dunjko, \& M. Olshanii,
Thermalization and its mechanism for generic isolated quantum systems,
Nature {\bf 452}, 854 (2008);\\
A. Laeuchli \& C. Kollath, 
Spreading of correlations and entanglement after a quench in the
one-dimensional Bose-Hubbard model, 
J. Stat. Mech. {\bf 2008}, P05018 (2008);  \\
P. Barmettler, A. M. Rey, E. Demler, M. D. Lukin, I. Bloch, \& V. Gritsev,
Quantum many-body dynamics of coupled double-well superlattices,
Phys. Rev. A {\bf 78}, 012330 (2008);\\ 
A. Flesch, M. Cramer, I.P. McCulloch, U. Schollwoeck, \& J. Eisert, 
Probing local relaxation of cold atoms in optical superlattices,
Phys. Rev. A {\bf 78}, 033608 (2008);\\
G. Roux, On quenches in non-integrable quantum many-body systems: the
one-dimensional Bose-Hubbard model revisited, Phys. Rev. A {\bf 79}, 021608(R) (2009);\\
P. Barmettler, M. Punk, V. Gritsev, E. Demler \& E. Altman, 
Relaxation of antiferromagnetic order in spin-1/2 chains following a quantum
quench, Phys. Rev. Lett. {\bf 102}, 130603 (2009);\\
S. R. Manmana, S. Wessel, R. M. Noack, \& A. Muramatsu,
Time evolution of correlations in strongly interacting fermions after a
quantum quench, Phys. Rev. B {\bf 79}, 155104 (2009).



\bibitem{cc-06}
P. Calabrese \& J. Cardy, 
Time-dependence of correlation functions following a quantum quench,
Phys. Rev. Lett. {\bf 96}, 136801 (2006);\\
P. Calabrese \& J. Cardy, 
Quantum quenches in extended systems, J. Stat. Mech. {\bf 2007}, P06008 (2007).

\bibitem{gdlp-07} 
V. Gritsev, E. Demler, M. Lukin, \& A. Polkovnikov,
Spectroscopy of collective excitations in interacting low-dimensional
many-body systems using quench dynamics,
Phys. Rev. Lett. {\bf 99}, 200404 (2007).

\bibitem{IS}
E. Barouch, B. McCoy, \& M. Dresden, 
Statistical mechanics of the XY model. I,
Phys. Rev. A {\bf 2}, 1075 (1970); \\
E. Barouch \& B. McCoy, 
Statistical mechanics of the XY model. III, 
Phys. Rev. A {\bf 3}, 3127 (1971);\\
F. Igloi \& H. Rieger, 
Long-Range correlations in the nonequilibrium quantum relaxation 
of a spin chain,
Phys. Rev. Lett. {\bf 85}, 3233 (2000);\\
K. Sengupta, S. Powell, \& S. Sachdev, 
Quench dynamics across quantum critical points,
Phys. Rev. A {\bf 69}, 053616 (2004); \\
P. Calabrese \& J. Cardy, 
Evolution of entanglement entropy in one dimensional systems,
J. Stat. Mech. {\bf 2005}, P04010 (2005);\\
R. W. Cherng \& L. S. Levitov, 
Entropy and correlation functions of a driven quantum spin chain,
Phys. Rev. A  {\bf 73}, 043614 (2006);\\
V. Mukherjee, U. Divakaran, A. Dutta, \& D. Sen,
Quenching dynamics of a quantum XY spin-1/2 chain in a transverse field,
Phys. Rev. B {\bf 76}, 174303 (2007);\\
V. Eisler \& I. Peschel, Entanglement in a periodic quench,
Ann. Phys. (Berlin) {\bf 17}, 410 (2008);\\
M. Fagotti \& P. Calabrese, 
Evolution of entanglement entropy following a quantum quench: Analytic results
for the XY chain in a transverse magnetic field, 
Phys. Rev. A {\bf 78}, 010306(R) (2008);\\
V. Mukherjee, A. Dutta, \& D. Sen, 
Defect generation in a spin-1/2 transverse XY chain under repeated quenching
of the transverse field, 
Phys. Rev. B {\bf 77}, 214427 (2008);\\
D. Rossini, A. Silva, G. Mussardo \& G. Santoro, 
Effective thermal dynamics following a quantum quench in a spin chain,
Phys. Rev. Lett. {\bf 102}, 127204 (2009);\\
A. Silva, The statistics of the work done on a quantum critical system by
quenching a control parameter, 
Phys. Rev. Lett. {\bf 101}, 120603 (2008).

\bibitem{fcc-09}
A. Faribault, P. Calabrese \& J.-S. Caux, Quantum quenches from integrability: the fermionic pairing model
J. Stat. Mech. {\bf 2009}, P03018  (2009).

\bibitem{bcs-57}
J. Bardeen, L. N. Cooper \& J. R. Schrieffer,
Microscopic theory of superconductivity,
Phys. Rev. {\bf 106}, 162 (1957);  \\
J. Bardeen, L. N. Cooper \& J. R. Schrieffer,
Theory of superconductivity, 
Phys. Rev. {\bf 108}, 1175 (1957).


\bibitem{anderson}
P. W. Anderson, Random-Phase Approximation in the Theory of Superconductivity,
Phys. Rev. {\bf 112}, 1900 (1958).

\bibitem{rs-62}
R. W. Richardson, 
A restricted class of exact eigenstates of the pairing-force Hamiltonian,
Phys. Lett. {\bf 3}, 277 (1963);\\
R. W. Richardson, Application to the exact theory of the pairing model to some
even isotopes of lead, 
Phys. Lett. {\bf 5}, 82 (1963); \\
R. W. Richardson \& N. Sherman, 
Exact eigenstates of the pairing-force Hamiltonian,
Nucl. Phys. {\bf 52}, 221 (1964);\\
R. W. Richardson \& N. Sherman, 
Pairing models of Pb$^{206}$, Pb$^{204}$ and Pb$^{202}$,
Nucl. Phys. {\bf 52}, 253 (1964).


 \bibitem{rev}
J. von Delft \& D. C. Ralph, Spectroscopy of discrete energy levels in
ultrasmall metallic grains,  
Phys. Rep. {\bf 345}, 61 (2001);\\
J. Dukelsky S. Pittel \& G. Sierra, 
Exactly solvable Richardson-Gaudin models for many-body quantum systems,
Rev. Mod. Phys. {\bf 76},  643 (2004).


\bibitem{camb}
M. C. Cambiaggio, A. M. F. Rivas, and M. Saraceno, Integrability of the pairing hamiltonian,
Nucl. Phys. A {\bf 624} (1997) 157.

\bibitem{BetheZP71}
H. Bethe,  Zur theorie der metalle, Z. Phys. {\bf 71}, 205 (1931).

\bibitem{zlmg-02}
H.-Q. Zhou J. Links, R.H. McKenzie, \& M.D. Gould,
Superconducting correlations in metallic nanoparticles: exact solution of the
BCS model by the algebraic Bethe ansatz, 
Phys. Rev. B {\bf 65}, 060502(R) (2002);\\
J. Links, H.-Q. Zhou, R.H. McKenzie, \& M.D. Gould, 
Algebraic Bethe ansatz method for the exact calculation of energy spectra and
form factors: applications to models of Bose-Einstein condensates and metallic
nanograins, 
J. Phys. A {\bf 36}, R63 (2003).

\bibitem{largealt}
E. A. Yuzbashyan, A. A. Baytin, \& B. L. Altshuler,
Strong coupling expansion for the pairing Hamiltonian,
Phys. Rev. B {\bf 68}, 214509 (2003). 




\bibitem{algo}
S. Rombouts, D. Van Neck and J. Dukelsky, Solving the Richardson equations for fermions, Phys. Rev. C  {\bf 69} , 061303 (2004);\\ 
F. Dominguez, C. Esebbag, and J. Dukelsky, Solving the Richardson equations close to the critical points,J. Phys. A {\bf 39}, 11349 (2006);\\
M. Sambataro, Pair condensation in a finite Fermi system,Phys. Rev. C {\bf 75}, 054314 (2007);\\
R. W. Richardson, Numerical study of the 8-32-particle eigenstates of the
pairing Hamiltonian, Phys. Rev. {\bf 141}, 949 (1966).

\bibitem{largeroman}
J. M. Roman, G. Sierra, \& J. Dukelsky,
Elementary excitations of the BCS model in the canonical ensemble, 
Phys. Rev. B {\bf 67}, 064510 (2003).

\bibitem{fcc-08}
A. Faribault, P. Calabrese, \& J.-S. Caux, 
Exact mesoscopic correlation functions of the pairing model,
Phys. Rev. B {\bf 77}, 064503 (2008).

\bibitem{s-89}
N. A. Slavnov, On scalar products in the algebraic Bethe ansatz,
Teor. Mat. Fiz. {\bf 79}, 232 (1989).


\bibitem{bl-06}
R. A. Barankov \& L. S. Levitov,
Synchronization in the BCS pairing dynamics as a critical phenomenon,
Phys. Rev. Lett. {\bf 96}, 230403 (2006).

\bibitem{bcs}
E. A. Yuzbashyan, B. L. Altshuler, V. B. Kuznetsov, \& V. Z. Enolskii,
Nonequilibrium Cooper pairing in the nonadiabatic regime,
Phys. Rev. B {\bf 72}, 220503(R) (2005);\\
E. A. Yuzbashyan \& M. Dzero, 
Dynamical vanishing of the order parameter in a fermionic condensate,
Phys. Rev. Lett. {\bf 96}, 230404 (2006); \\
E. A. Yuzbashyan, O. Tsyplyatyev, \& B. L. Altshuler,
Relaxation and persistent oscillations of the order parameter in the
non-stationary BCS theory, 
Phys. Rev. Lett. {\bf 96}, 097005 (2006);\\
M. Dzero, E. A. Yuzbashyan, B. L. Altshuler, \& P. Coleman, 
Spectroscopic signatures of nonequilibrium pairing in atomic Fermi gases,
Phys. Rev. Lett. {\bf 99}, 160402 (2007); \\
R. A. Barankov \& L. S. Levitov, 
Excitation of the dissipationless Higgs mode in a fermionic condensate,
0704.1292;\\
A. Tomadin, M. Polini, M. P. Tosi, \& R. Fazio,
Nonequilibrium pairing instability in ultracold Fermi gases with population
imbalance, Phys. Rev. A {\bf 77}, 033605 (2008).

\bibitem{ao-02}
A. Mastellone, G. Falci, \& R. Fazio,
A small superconducting grain in the canonical ensemble,
Phys. Rev. Lett. {\bf 80}, 4542  (1998);\\
L. Amico \& A. Osterloh, 
Exact correlation functions of the BCS model in the canonical ensemble,
Phys. Rev. Lett. {\bf 88}, 127003 (2002).

\bibitem{cm-05}
J.-S. Caux \& J.-M. Maillet, 
Computation of dynamical correlation functions of Heisenberg chains in a field,
Phys. Rev. Lett. {\bf 95}, 077201 (2005);\\
J.-S. Caux, R. Hagemans \& J.-M. Maillet,
Computation of dynamical correlation functions of Heisenberg chains: 
the gapless anisotropic regime,
J. Stat. Mech. {\bf 2005},  P09003 (2005).

\bibitem{cc-06b}
J.-S. Caux \& P. Calabrese,
Dynamical density-density correlations in the one-dimensional Bose gas,
Phys. Rev. A {\bf 74}, 031605 (2006);\\
J.-S. Caux, P. Calabrese \& N. A. Slavnov,
One-particle dynamical correlations in the one-dimensional Bose gas,
J. Stat. Mech. {\bf 2007}, P01008 (2007).





\end{thebibliography}

\end{document}